\pdfoutput=1

\documentclass[aps,prd,reprint,groupedaddress, longbibliography]{revtex4-2}
\usepackage{pgfplots}
\pgfplotsset{compat=newest}
\usepackage{tikz}
\usepackage[left]{lineno}
\usepackage{longtable}
\usepackage{graphicx}
\usepackage{amssymb}
\usepackage{fancyvrb}
\usepackage{verbatimbox}
\usepackage{listings}
\usepackage{array}
\usepackage{longtable}
\usepackage{amsmath}
\usepackage{amssymb}
\usepackage{enumitem}
\usepackage[title]{appendix}
\usepackage{nicefrac,xfrac}
\usepackage{blindtext}
\begin{document}


\newcolumntype{P}[1]{>{\centering\arraybackslash}p{#1}}

\title{Ionization efficiency for nuclear recoils in silicon from $\sim 50$ eV to $3$ MeV}

\author{Y. Sarkis, A. Aguilar-Arevalo and J.C. D'Olivo}

\affiliation{Instituto de Ciencias Nucleares, Universidad Nacional Aut\'onoma de M\'exico,  04510 CDMX, Mexico}

\date{\today}

\begin{abstract}
We present a model for the nuclear recoil ionization efficiency in silicon based on an extension of Lindhard’s theory where atomic bond disruption is modeled as a function of the initial ion energy, the interatomic potential, and the average  ion-vacancy production energy. A better description of the electronic stopping than the one assumed by Lindhard, the effect of electronic straggling, as well as charge screening and Coulomb repulsion effects of ions are also considered. The model describes the available data over nearly four orders of magnitude in nuclear recoil energy.

\end{abstract}


\maketitle

 \textbf{\textit{Introduction.}} Silicon is a common detection medium in experiments searching for the low-energy depositions expected from coherent elastic neutrino-nucleus scattering (CE$\nu$NS) or dark matter (DM) interactions. In these applications, Si detectors typically observe the ionization signal produced by a nuclear recoil after  interaction with the incident particle. The ionization efficiency, or quenching factor (QF), defined  as the ratio of the amount of electronic excitation produced by a nuclear recoil to that produced by a recoiling electron of the same energy, is a key feature of these detectors. As early as 1963, Lindhard \cite{lindhard:1963} crafted the basic integral equation describing the partitioning of the energy deposited by a nuclear recoil between electronic and atomic motion. Since then, his approximate solution for the QF has been known to describe the data \cite{Brian,SattlerDat,Zech} at high energies  ($>10$ keV), where atomic binding energy effects can be safely ignored. It has been shown \cite{QFUNAM} that when properly introduced into the model, a constant average binding energy of the order of 150 eV, consistent with that associated with $2s2p$ shell electrons in Si, can describe the data available at the lowest energies \cite{chavarria:2016,izraelevitch:2017}.  Although a constant binding energy model works well for the existing data, it predicts a cut-off at $\sim$300 eV, significantly greater than the $36$~eV of energy required in average to create a stable ion-vacancy pair (Frenkel \cite{Frenkel1926} pair), which can be argued to represent an upper bound on the minimum energy required to be deposited by a nuclear recoil to produce an observable ionization signal in a Si detector.

In order to obtain a model that is valid at lower nuclear recoil energies, in this work we implement several improvements to the integro-differential equation with binding energy presented in \cite{QFUNAM}. 
 We assume that ionization can only occur when an ion can move freely through the lattice but, instead of a constant threshold, the energy lost due atomic bond breaking  
 is  described as a ``binding energy" function dependent on the energy. In the low energy regime it corresponds to the Frenkel pair creation energy, while at higher energies it describes the inner atomic excitation energies. We also make use of improved modeling of the electronic stopping power at low energies, and add a term related to electronic straggling \cite{Lindhard1954,PeterSigmundV1} to the equation.
 

Consideration of the aforementioned effects  results in a first-principles model of the QF capable of describing the available published data in Si from 0.68 keV to 3 MeV nuclear recoil energy, and provides a prediction for this quantity down to a few tens of eV.

 \textbf{\textit{The nuclear recoil QF and Lindhard's basic integral equation}.} 
When a neutrino or a DM particle elastically scatters off a nucleus in a Si detector, the struck ion recoils with an energy $E_R$. If the timescale of the collision is much smaller than that of atomic processes, the ion will loose some energy to the atomic degrees of freedom emerging from the collision with kinetic energy $E=E_R-U$, where $U$ is the energy used to disrupt the atomic binding. In general, $U$ is not
limited to the energy needed to remove the ion from its site, but it can also include contributions to excitation or ionization of bound atomic electrons, and therefore effectively incorporates the Migdal effect \cite{migdal1,migdal2}.

Lindhard's theory \cite{lindhard:1963} assumes that $E_R$, the total energy deposited by the incoming particle in the material, is separated between the energy given to electrons, $H$ and the energy given to atomic motion, $N$ so that $E_R=H+N$. For binary collisions, the Ziegler scaling length \cite{Ziegler} $a=(0.8853)a_0/Z$, with $Z=Z_1^{0.23}+Z_2^{0.23}$, and $a_0$ the Bohr radius, can be used with suitable values of the effective atomic numbers of the incident ion and the target atom $Z_1$ and $Z_2$, respectively, to define the reduced dimensionless quantities $\varepsilon=CE$, $\eta=CH$, and $\nu=CN$ where $C=16.2616/Z Z_1 Z_2$ keV$^{-1}$. 
Assuming that this separation holds on average over the large number of collisions occurring in the process, the average quantities also satisfy $\varepsilon_R=\bar\eta+\bar\nu$. The nuclear recoil QF is given by $f_n=\bar\eta/\varepsilon_R=(\varepsilon+u-\bar\nu)/(\varepsilon+u)$,where $\varepsilon_R=\varepsilon+u$, and $u=CU$.

Consider an ion moving in the material with energy $E$ and colliding with an atom, transferring an energy $T_n$ to its center of mass and an energy $\sum_i T_{ei}$ to a certain number of ionized electrons. The ion will scatter with an energy $E-T_n-\sum_i T_{ei}$, and the struck atom will recoil with an energy $T_n-U$.
Lindhard's basic integral equation for $\bar\nu(\varepsilon)$,
\begin{equation}\label{Eq:BasIntEq}
\int d\sigma_{n,e}\left[  \bar\nu (E-T_{n}-\Sigma_i T_{ei}) +
\bar\nu (T_{n}-U) -\bar\nu (E) \right] =0,
\end{equation}
states that the average energy given to atomic motion by the initial ion with energy $E$ equals the sum of the contributions 
of the scattered ion and the struck recoiling ion, where the contribution due to ejected electrons has been neglected  (approximation I). Integration over the total nuclear and electronic cross sections $\int d\sigma_{n,e}$ represents the sum over all possible impact parameters for nuclear and electronic collisions.

Lindhard made use of four additional approximations to construct an approximate integro differential equation for $\bar\nu(\varepsilon)$: II) the energy transferred to ionized electrons is small; III) electronic and atomic collisions can be treated separately; IV) $T_n$ is also small compared to $E$; V) neglect the binding energy $U$. As has been shown in \cite{QFUNAM}, dropping V results in a higher order approximation to the integro-differential equation. In what follows, we adopt the same approach.

\textbf{\textit{Improved integro-differential equation}}
Using approximations I-IV, Eq.(\ref{Eq:BasIntEq}) can be written in a form suitable for numerical solution, where the electronic stopping and straggling appear naturally.
If $\Delta E$ is the energy lost by an ionizing projectile, straggling is defined as the mean-square fluctuation $\Omega^{2}= \langle(\Delta E- \langle\Delta E\rangle)^2 \rangle$ \cite{PeterSigmundV1}.
Lindhard's partitioning of the deposited energy into electronic and atomic motion, guarantees that the electronic and nuclear contributions to straggling are equal \cite{lindhard:1963}.
The electronic straggling per unit length can be expressed as $N_e^{-1}(d\Omega^2/dR)_e = \int d\sigma_e(\Sigma_i T_{ei})^2$ \cite{PeterSigmundV1}, where $N_e$ is the electron number density and $R$ is the distance traveled by the projectile. In terms of the reduced quantities $\varepsilon$, $\omega^2=C^2\Omega^2$ and $\rho=\pi a^2N_eR$, we have  $W(\varepsilon) = d\omega^2/d\rho$.
Similarly, for the electronic stopping power  $N_e^{-1}(dE/dR)_e=\int d\sigma_e(\Sigma_i T_{ei})$, and $S_e(\varepsilon)=d\varepsilon/d\rho$.

Expanding the first term in Eq.(\ref{Eq:BasIntEq}) up to second order in $\Sigma_i T_{ei}/(E-T_n)$ (approximations II-IV), as
$\bar{\nu}\left(E-T_{n}-\Sigma_{i} T_{e i}\right)  \approx  
 \bar{\nu}\left(E-T_{n}\right)-\bar{\nu}^{\prime}(E)\left(\Sigma_{i} T_{e i}\right)
 +\bar{\nu}^{\prime \prime}(E) T_{n}\left(\Sigma_{i} T_{e i}\right) +\frac{1}{2}\bar\nu''(E){\left(\Sigma_{i} T_{e i}\right)^{2}}$,
  and following the arguments in \cite{QFUNAM}, the electronic stopping power and straggling can be identified in the terms containing $\Sigma_i  T_{ei}$ to first and second order, respectively.
With these considerations, we write the modified simplified integral equation with straggling as
\begin{equation}\label{Ec:MainintegroDifStraggling}
\begin{aligned}
-\frac{1}{2} & \varepsilon S_e (\varepsilon)\left(1 +\frac{W(\varepsilon)}{S_e (\varepsilon) \varepsilon} \right) \bar{\nu}^{\prime \prime}(\varepsilon)+S_e (\varepsilon)\bar{\nu}^{\prime}(\varepsilon)=\\
&\int_{\varepsilon u}^{\varepsilon^{2}} d t \frac{f\left(t^{1 / 2}\right)}{2 t^{3 / 2}}  \times[\bar{\nu}(\varepsilon-t / \varepsilon)+\bar{\nu}(t / \varepsilon-u)-\bar{\nu}(\varepsilon)],
\end{aligned}
\end{equation}
where $u(\varepsilon)$ is the binding energy, $t=\varepsilon^{2}\sin^{2}(\theta/2)$, with $\theta$ the scattering angle in the center of mass frame, and $f(t^{1/2})$ is related to the nuclear differential cross section in the Lindhard-Scharff-Schiott (LSS) approximation,
$d\sigma_n= \pi a^2 dt f(t^{1/2})/2t^{3/2}$
\cite{lindhard:1963}. 

The contribution of $S_e(\varepsilon)$ in Eq.~(\ref{Ec:MainintegroDifStraggling}) is dominant  compared to  $W(\varepsilon)$. For the latter we use the parametrization given by Wilson \textit{et al.} \cite{WilsonPot}, 
$W(\varepsilon) = 1/[4+A\varepsilon^{-B} + C\varepsilon^{-D}]$, where the constants $A, B, C$, and $D$ depend on the choice of the interatomic potential, characterized by the choice of the screening function $\phi(x)$. In this work, we consider four such potentials: Tomas Fermi, Moli\`ere, and Average from Ref. \cite{WilsonPot}, and Ziegler from \cite{Ziegler}, each giving a model for $u(\varepsilon)$ and  $f(t^{1/2})$.
For $S_e(\varepsilon)$ we consider three detailed models, which discussed below.

\textbf{\textit{Electronic stopping power ($S_e$)}.}
 Lindhard's theory of the electronic stopping power \cite{lindhard1961} is appropriate for atomic collisions down to energies of the order of a few keV. It assumes point-like interactions between an incident atom with $Z$ electrons, whose velocity is not affected by interatomic potentials,
 and the electron cloud of the target atom, which is approximated by a Fermi gas. An important assumption in the theory is that all the electronic states of the degenerate gas with energies up to the Fermi energy $E_F$ of the system, participate in the ionization process.
 %
 %
 Lindhard also disregards Coulomb repulsion effects between colliding atoms, which at low energies ($<1$ keV) prevent electron clouds from penetrating each other completely, allowing only interactions in regions of low electron density \cite{CoulombRepultionEffect}.
 When taken into account, the Coulomb repulsion effects lead to an appreciable departure from the proportionality with velocity of the electronic stopping power assumed by Lindhard.

\begin{figure}
    \centering
    \includegraphics[scale=0.45]{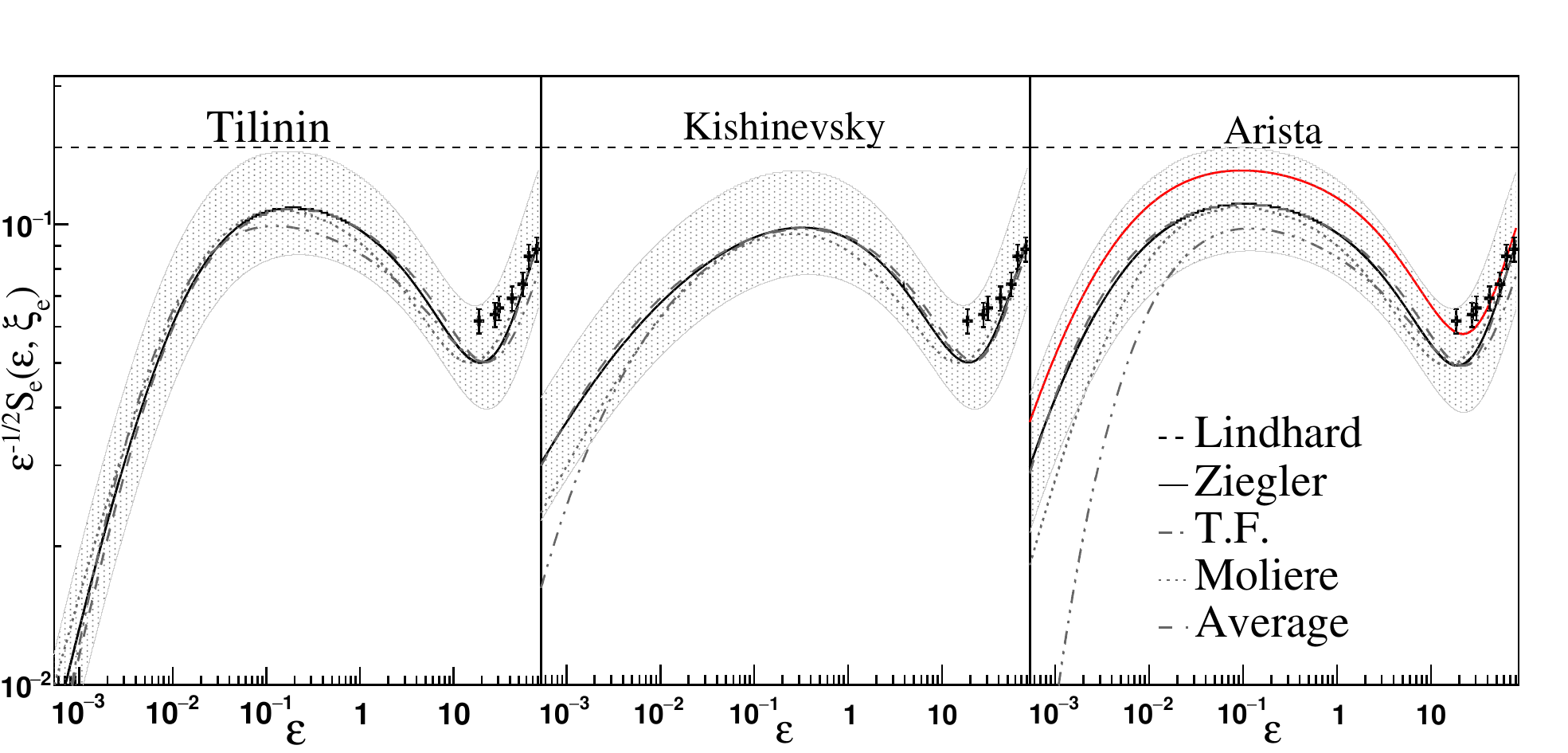}
    \caption[Cap]{Electronic stopping power $S_e(\varepsilon, \xi_e)$, with $\xi_e=1.26$, from Tilinin, Kishinievsky, and Arista, for the various interatomic potentials: Ziegler (solid), TF (dash-dot), Moliere (dotted), Average (dash-dot-dot). The gray band shows the range of $\xi_e$ consistent with the spread of QF measurements in Fig.~\ref{fig:QFResult}. Available data for Si-on-Si \cite{SeData1-ARSTILA2000473,SeData2-N.Hecking,SeData3-PhysRevB.68.235317} are shown as points with error bars . The red line in the right panel is the Arista-Ziegler combination with $\xi_e=1.46$, which best fits the data. The Lindhard stopping power ($\propto ~\varepsilon^{1/2}$) is shown as the constant dashed line. }
    \label{fig:Tilinin_Se}
\end{figure}

In general, a detailed model of electronic stopping requires a non-perturbative analysis of electron dynamics \cite{Nonperturbative}. 
When the moving ion collides with an atom in the lattice, multiple inner electronic transitions and electron promotion will occur, contributing to electronic stopping even for low sub-keV energies \cite{Ufano}.
Two general approaches are considered in the literature to study electronic stopping: kinetic theory \cite{SemiclassicalCrossSectionDer,SemiclassicalCrossSectionDer2} and the dielectric function approach, pioneered by Lindhard \cite{lindhard:1961}.
From the first category, in this work we consider the models due to Tilinin \cite{Tilinin,KineticTheorySe_Der} and Kishinevsky \cite{Kishi,Firsov,Kpar1,Kpar2,Kpar3}, and from the second, the model due to Arista \cite{Arista, PeterSigmundV2}. All three are theoretically well founded semi-classical models, which assume that the kinetic energy of the incoming ion is always sufficient to effectively excite an atom of the material by electron promotion effects. They also rely on the determination of the distance of closest approach of the ion in the interatomic potential. The three models are compared in Fig. \ref{fig:Tilinin_Se} for values of $\varepsilon$ between $0.0006$ and $80$ (about 20 eV to 3 MeV for Si). Details on our implementation of these models can be found in the supplemental materials.

{\it Scaling length.}
A general feature of electronic stopping studies is the use of the Thomas Fermi (TF) model to define the scaling length $a$, introduced above. All the atomic electrons in the free gas, with average kinetic energy $3/5 E_F$, are assumed to participate in the momentum transfer, an approximation valid only for sufficiently high energies. However, as noted in \cite{Tilinin}, at low energies, the excitation energies of the electrons are typically much smaller than $E_F$ and, since transitions between occupied states are forbidden, the only electrons participating in the momentum transfer are those close to the Fermi level, with kinetic energy $\sim E_F$. This results in a correction to the TF scaling length by a factor of $5/3$ at low energies. 
For models like those of Refs.\cite{Tilinin}, \cite{Kishi}, and \cite{Arista}, where $S_e \propto  \varepsilon^{1/2} \tau(\varepsilon,Z_1,Z_2)$, with $\tau$ being a dimensionless function incorporating the effects of Coulomb repulsion, it can be shown that changing the scaling length introduces a change in $S_e$ dominated by a multiplicative factor of $\xi_e=(5/3)^{1.5}\approx 2.152$.
For high energies, in principle, $\xi_e=1$. Averaging the low and high energy effects on the scaling $a$, an effective value of $\xi_e$ between 2.152 and 1 may be used for all energies.
Notice, however, that Lindhard introduced by hand \cite{Sigmund2008Resume} a semi-empirical factor of $\approx Z^{1/6}$ (1.55 for Si) to match the stopping power data available at his time, which could be interpreted as the effective value of $\xi_e$ suitable for high energies.

%

A change in $a$ will also change the interatomic potential, by a factor of $\xi_e^{-2/3}$, and, therefore, introduce  modifications in any quantity depending on it, such as the nuclear and electronic stopping, and the variable binding energy in the model discussed below.
In solving Eq.(\ref{Ec:MainintegroDifStraggling}) we will take into account the uncertainty in the value of $\xi_e$, considering it as an additional parameter of the electronic stopping, {\it i.e.} changing $S_e(\varepsilon)$ to $S_e(\varepsilon, \xi_e)$.

{\textit{Bohr stripping and Z oscillations.}} When a projectile of velocity $v$ collides with a target atom, accoring to Bohr's  criterion \cite{BohrStrip1}, the projectile is stripped of those electrons with orbital velocities smaller than $v$. This leads to ions having an effective number of electrons $Z^{\dagger}<Z_{\alpha}$ ($\alpha=1,2$), which can be computed using the condition $v\leq \sqrt{3}v_F$ \cite{BohrStrip2}. In terms of the screening function $\phi(x)$, $Z^{\dagger}=Z_{\alpha}[1- \phi(x_0) +x_0\phi'(x_0)]$, where $x_0$ is the distance at which $v=\sqrt{3} v_F(x_0)$.
Friedel \cite{Friedel} predicted that for low energies ($\ll$ 30 MeV), the change in electron density around an ion or impurity induces an oscillatory behavior of the electron transport cross section as a function of $Z^\dagger$, which is inherited by the electronic stopping power of the material.
For Si we used the oscillatory factor computed in \cite{ZOscilations} normalized to its value at $Z_1=12$ and applied it as an energy dependent multiplicative correction to the electronic stopping. For each ion energy an effective $Z^\dagger$ is extracted from Bohr's criterion. We notice that for Si, including this effect results in good agreement with measurements of the electronic stopping power in the interval from 680 eV to 3 MeV, with effective values of $\xi_e$ ranging between 1.3 and 1.7, consistent with Lindhard's observed value, see Fig.~\ref{fig:Tilinin_Se}.

\begin{figure}
    \hspace{-0.3cm}
    \includegraphics[scale=0.48]{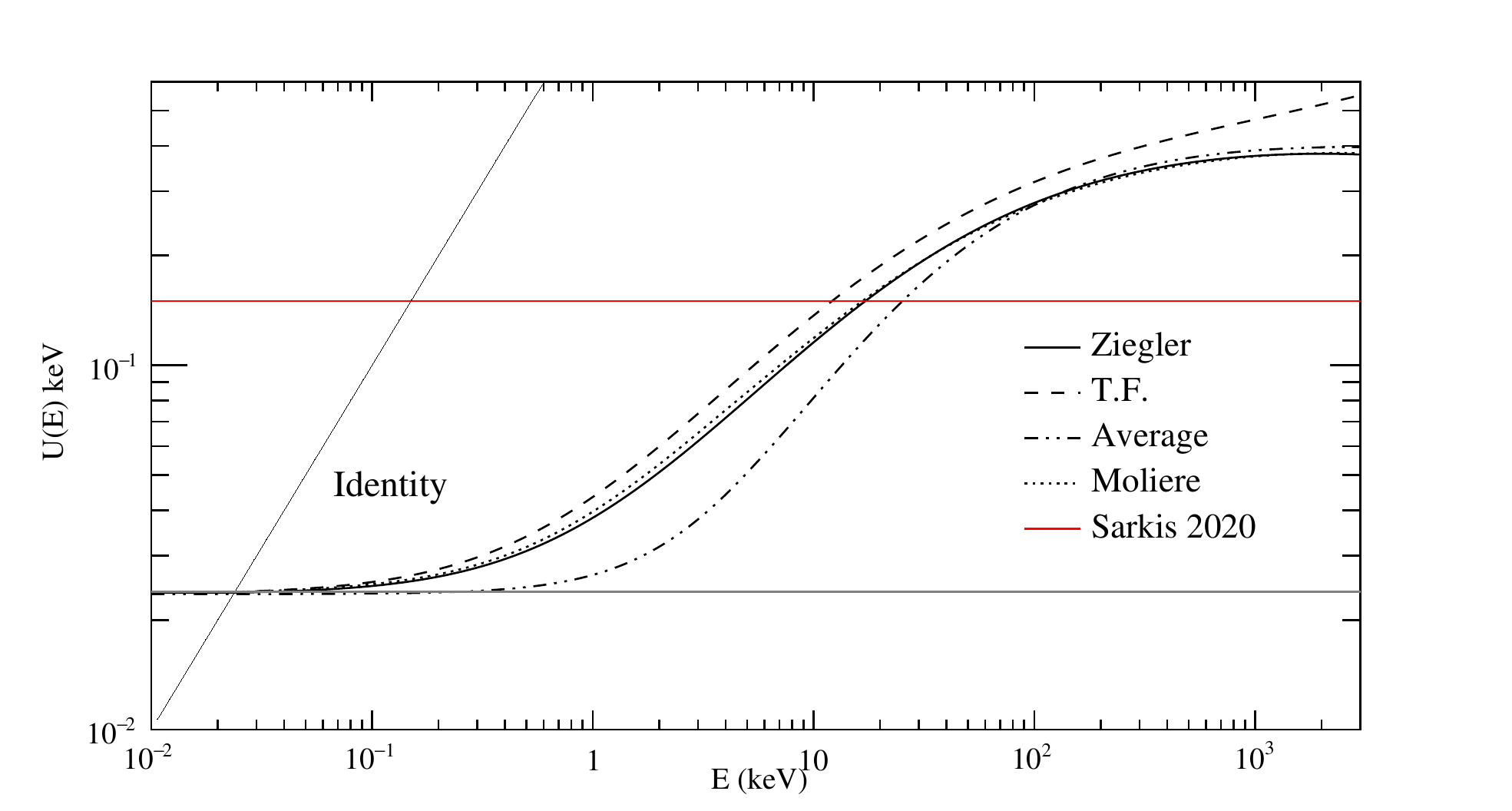}
    \caption{Variable binding energy model with $\xi_e =1.26$, based on EDFT, see Eq.(\ref{Eq:bindingEModel}) , for the four interatomic potencials. Includes both, the Frenkel-pair creation energy in Si and the inner electron excitations. }
    \label{fig:Umodel}
\end{figure}

\textbf{\textit{Binding energy model}.}
Part of the energy transferred to a target atom in the material by an incident ion is used to excite atomic degrees of freedom and extract the atom from its site in the lattice. The energy lost in these processes, $U$ in Eq.(\ref{Eq:BasIntEq}), is in general dependent on the incident ion energy, $E$. At low energies it corresponds to the energy required to remove the atom from a lattice site creating a Frenkel-pair (FP) \cite{Frenkel1926}, while at higher energies it is dominated by the energy spent in disrupting the atomic binding in the outer valence orbitals.
To account for both effects, we define an energy dependent binding energy function $U(E)=U_{\rm{FP}}+ U_{\rm{EDFT}}(E)$, shown in Fig. (\ref{fig:Umodel}). 
The first constant term is the FP creation energy, for which we consider the interval $U_{\rm{FP}}=23.6^{+15.2}_{-12.1}$ eV, where the central value is consistent with the weighted average of the  predictions from H\"olstrom, Stallinger-Weber, and Tersov quoted in \cite{FrenkelSi}.The lower and upper limits cover the wide variability of both, experimental measurements and predictions, including the highest range reported from recent simulations based on electron density functional theory (EDFT). 
For the second term of $U(E)$, the energy stored in the electron cloud of an atom in the EDFT model, from the distance of closest approach, $x(E,\xi_e)$, to infinity, with a given choice for the interatomic potential,  is considered a measure of the energy absorbed by the struck atom prior to its recoil in the sudden approximation. This energy is computed in terms of the screening function $\phi$  as 

\begin{equation}\label{Eq:bindingEModel}
U_{\rm{EDFT}}(E,\xi_e)={U_{0}} \int^{\infty}_{x} dy \: \phi^{5/2}(y,\xi_e)/ y^{1/2}    
\end{equation}
 where ${U_{0}}= 18.6/\xi^{4/3}_e \; Z^{1.23}$~eV.\\

\textbf{\textit{Solving the equation with variable $u(\varepsilon)$}.} 
Generalizing the ideas in \cite{QFUNAM} for an energy dependent $u(\varepsilon)$, it can be seen that Eq.~(\ref{Ec:MainintegroDifStraggling}) is only applicable for $\varepsilon \geq u(\varepsilon)$. When the equality holds, the rhs of Eq.(\ref{Ec:MainintegroDifStraggling}) is equal to zero, since the upper and lower limits of integration are equal. This will occur at the threshold energy $\varepsilon^*=u(\varepsilon^*)$, which for silicon is only slightly greater than the FP creation energy.
The appropriate generalization of the parameterized solution for $\bar\nu(\varepsilon$) studied in \cite{QFUNAM} is the following:
\begin{equation}
\label{eq:solparam}
\bar\nu[s(\varepsilon)] = 
\left\lbrace
\begin{array}{ll}
s(\varepsilon) + u(\varepsilon), & s(\varepsilon) \leq u(\varepsilon), \\
s(\varepsilon) + u(\varepsilon) - \lambda[s(\varepsilon)], & s(\varepsilon)  \geq u(\varepsilon), \\
\end{array}
\right.    
\end{equation}

\noindent
where $s(\varepsilon)$ is any well behaved function of $\varepsilon$, such as those appearing as arguments of $\bar\nu$ in the rhs of Eq.(\ref{Ec:MainintegroDifStraggling}) and  $\lambda$ is a continuous function satisfying  $\lambda(\varepsilon^*)=0$, but with discontinuous derivatives at this point. This form will guarantee that at $\varepsilon^*$ the integrand in the rhs of Eq.(\ref{Ec:MainintegroDifStraggling}) goes to zero, as is required by energy conservation.
In the case $s(\varepsilon)=\varepsilon$, the discussion in \cite{QFUNAM} leading to the implementation of the shooting method for solving the equation for $\bar\nu(\varepsilon)$, applies for varying $u(\varepsilon)$ as well, with minor modifications.
For values below $\varepsilon^*$ it suffices that $\bar\nu(\varepsilon)$ take the form $\bar\nu(\varepsilon)=\varepsilon + u(\varepsilon)= \varepsilon_R$.
For $\varepsilon\geq \varepsilon^*$ the solution is found numerically.
To take into account the effect that the moving ion will lose some of its valence electrons as a consequence of its motion through the lattice, following \cite{CrystalModelYing_Tai_2002} we consider that the effective number of valence electrons in the incident ion is $Z_1=12$, while $Z_2=14$. Given that the Bohr Stripping effect describes the experimental data of electronic stopping power up to energies of the order of 3 MeV \cite{MagCEVNS2021}, we evaluate the solution from $U_{FP}$ to 3 MeV, in order to compare to QF measurements available at those energies (Sattler \cite {SattlerDat}).

\begin{figure}
    \centering
    \hspace{-0.5cm}
    \includegraphics[scale=0.48]{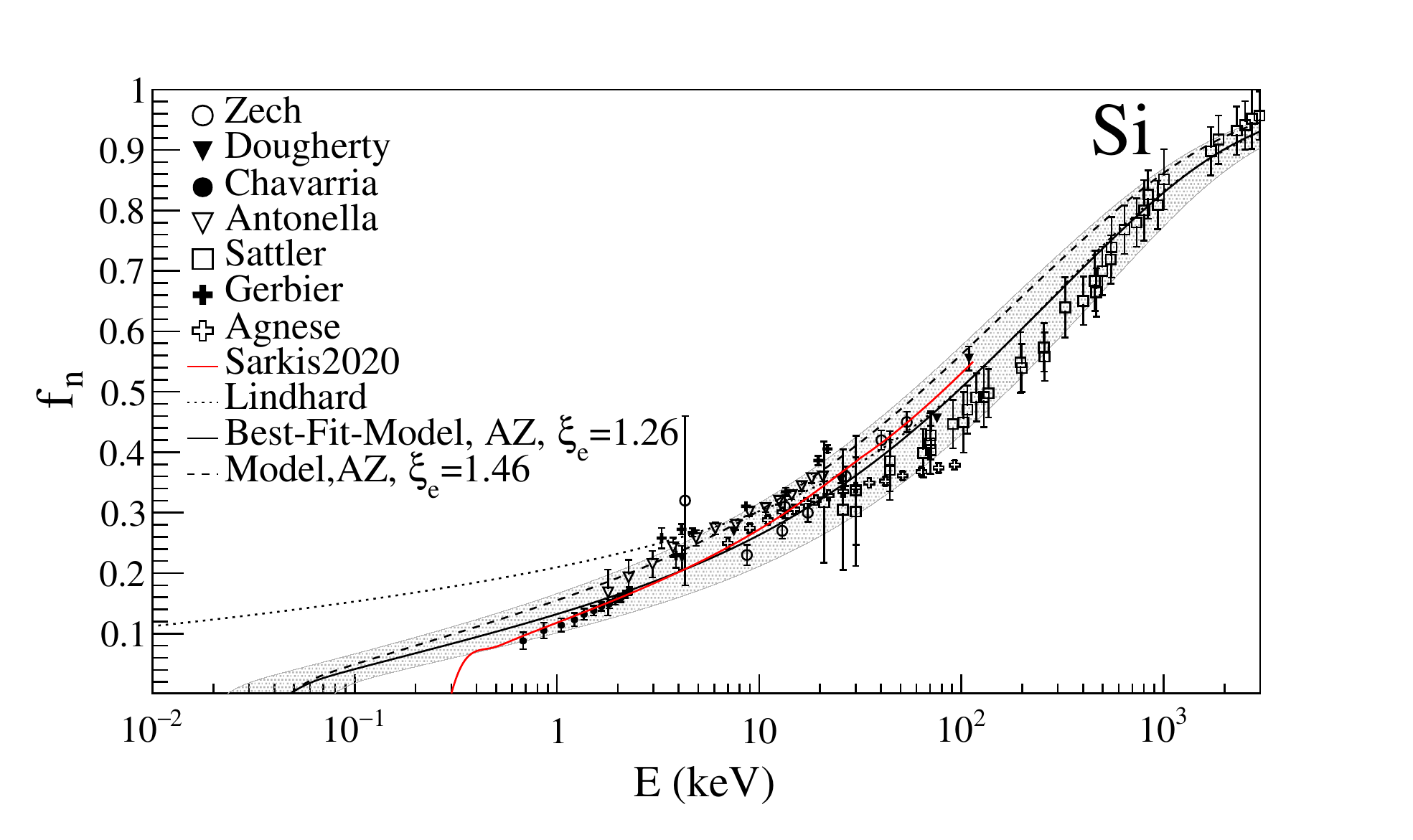}
    \caption{ Published Si QF data (points with error bars) compared to the Arista-Ziegler model with $\xi_e=1.26\pm0.25$ and $U_{FP}=23.6^{+15.2}_{-12.1}$~eV (black solid line with gray band). Also shown are the Arista-Ziegler model with $\xi_e=1.46$ (black dashed line), which best fits the $S_e$ data in Fig.\ref{fig:Tilinin_Se}, and the Lindhard model (dotted line).}
    \label{fig:QFResult}
\end{figure}

\textbf{\textit{Results.}}
We find the numerical solution to Eq.(\ref{Ec:MainintegroDifStraggling}) for the three different models of the electronic stopping power and the four interatomic potentials introduced. For each combination we evaluate the solution in a grid of 
values of $U_{FP}$ and $\xi_e$ constrained to lie within the physically motivated intervals discussed earlier. 
Each curve was compared to the available QF data in Si summarized in
Table II of \cite{QFUNAM}, to which we have added the Sattler \cite{SattlerDat}, Gerbier \cite{Gerbier} and Agnese \cite{AGNESE201871} data points. Overall, the data ($N_{data}=102$ points) span a range of energies from 680~eV to 3~MeV. 
We calculated the $\chi^2$ for each curve as $\chi^2=\sum_{i=1}^{N_{data}} (D_i-C_i)^2/\sigma_i^2$, where $D_i$ and $\sigma_i$ are the QF value and its error for data point $i$, and $C_i$ is the value predicted by the curve at the same energy.
Fixing the Frenkel pair energy at $U_{FP}=23.6$~eV, and varying $\xi_e$, we obtain that the model that best fits the data, shown as the black solid line in Fig.~\ref{fig:QFResult} corresponds to the combination of the Arista stopping power with Ziegler potential (Arista-Ziegler), with $\xi_e=1.26$ ($\chi^2/d.o.f=1870/101$). The high $\chi^2$ values are expected from the tension among the different data sets. We set an uncertainty on $\xi_e$ of $\pm0.25$ ($\Delta\chi^2\gtrsim 3100$) so as to approximately span the spread in the QF measurements.
Notice that the electronic stopping data in Si at higher energies prefer a larger value of $\xi_e$, around 1.46, but are consistent within the uncertainty interval determined for the QF measurements (see Fig. \ref{fig:Tilinin_Se}). The $\chi^2$ has a weak dependence on $U_{FP}$, and the solution at the center of the physically motivated interval has nearly the same goodness of fit than that at the edges ($\Delta \chi^2 \approx 40$). When $U_{FP}$ is allowed to vary freely, the fit prefers the lower edge of the interval.
Although the main effect of $U_{FP}$ is to set the threshold, it also has a small effect in the detailed shape of the curve at intermediate energies (1-100 keV). 

In Fig.~\ref{fig:QFResult} we show the Arista-Ziegler model with $\xi_e=1.26$ and $U_{FP}=23.6$~eV (black solid line). The gray band is drawn varying $\xi_e$ between 1.0 and 1.5, and varying $U_{FP}$ in the interval 11.5-38.8 eV. For comparison, we also show the model with $\xi_e=1.46$ (dashed black line), which gives the best fit to the $S_e$ for Si-on-Si data \cite{SeData1-ARSTILA2000473,SeData2-N.Hecking,SeData3-PhysRevB.68.235317} in Fig.~\ref{fig:Tilinin_Se}, for the Arista-Ziegler combination. The Sarkis 2020 \cite{QFUNAM} model (constant binding energy) is shown in the red line.


%
%
%
    In the supplemental materials we provide tabulated versions of the
    curves in Fig. \ref{fig:QFResult}, as well as a comparison of the curves obtained
    with fixed $\xi_e$ and $U_{FP}$ for the twelve combinations of models
    of the stopping power and interatomic potential.

\textbf{\textit{Conclusions.}}
\label{sec:conclusions}
In this letter, we present new results on the study of the QF for low-energy nuclear recoils in Si that introduce significant improvements over previous work based on the
Lindhard integral equation for the energy given to atomic motion \cite{QFUNAM}. We use more accurate descriptions of the electronic stopping power with several interatomic potentials. Instead of a constant binding energy, we treat it as a function of the initial ion energy, dependent also on the choice of the interatomic potential.
At low energies, this function is taken to correspond with the energy required to create a stable Frenkel pair.  
The high energy effects of Bohr stripping and Friedel Z-Oscillations were also incorporated and provide, simultaneously, a good match to the QF data at higher energies, also constraining the low energy behavior.
The main uncertainty on the models comes from the scaling parameter chosen to define the dimensionless energy scale $\xi_e$, affecting both, the low and high energy regimes. The value of the Frenkel pair creation energy plays a significant role at the lowest energies and might be more relevant when forthcoming data becomes available.
Slight tension between the sparse measurements of the Si-on-Si electronic stopping power at high energies and the QF data might be resolved with more measurements performed in both the low and high energy regimes.
Our model could be improved allowing the scaling factor to change with the ion energy. Phonon excitations, not included here, could also have an effect near the threshold. With these improvements, and  if more data of the electronic stopping becomes available it could be used in a combined fit with the QF data.
The found solution is able to describe reasonably well all the available published data for the nuclear recoil QF in Si from 50 eV to 3 MeV of nuclear recoil energy, and provides a unified theoretical picture of the effects of electronic and nuclear stopping from high energies down to the Frenkel pair creation threshold on the ionization efficiency in Si. 





\textbf{\textit{Acknowledgements}} This research was supported in part by DGAPA-UNAM grants number PAPIIT-IN106322 and PAPIIT-IT100420, and Consejo Nacional de Ciencia y Tecnolog\'ia (CONACYT) through grant CB-2014/240666. 

\begin{appendices}

\end{appendices}

\newpage
\bibliography{bibliografy}

\begin{thebibliography}{43}%
\makeatletter
\providecommand \@ifxundefined [1]{%
 \@ifx{#1\undefined}
}%
\providecommand \@ifnum [1]{%
 \ifnum #1\expandafter \@firstoftwo
 \else \expandafter \@secondoftwo
 \fi
}%
\providecommand \@ifx [1]{%
 \ifx #1\expandafter \@firstoftwo
 \else \expandafter \@secondoftwo
 \fi
}%
\providecommand \natexlab [1]{#1}%
\providecommand \enquote  [1]{``#1''}%
\providecommand \bibnamefont  [1]{#1}%
\providecommand \bibfnamefont [1]{#1}%
\providecommand \citenamefont [1]{#1}%
\providecommand \href@noop [0]{\@secondoftwo}%
\providecommand \href [0]{\begingroup \@sanitize@url \@href}%
\providecommand \@href[1]{\@@startlink{#1}\@@href}%
\providecommand \@@href[1]{\endgroup#1\@@endlink}%
\providecommand \@sanitize@url [0]{\catcode `\\12\catcode `\$12\catcode
  `\&12\catcode `\#12\catcode `\^12\catcode `\_12\catcode `\%12\relax}%
\providecommand \@@startlink[1]{}%
\providecommand \@@endlink[0]{}%
\providecommand \url  [0]{\begingroup\@sanitize@url \@url }%
\providecommand \@url [1]{\endgroup\@href {#1}{\urlprefix }}%
\providecommand \urlprefix  [0]{URL }%
\providecommand \Eprint [0]{\href }%
\providecommand \doibase [0]{https://doi.org/}%
\providecommand \selectlanguage [0]{\@gobble}%
\providecommand \bibinfo  [0]{\@secondoftwo}%
\providecommand \bibfield  [0]{\@secondoftwo}%
\providecommand \translation [1]{[#1]}%
\providecommand \BibitemOpen [0]{}%
\providecommand \bibitemStop [0]{}%
\providecommand \bibitemNoStop [0]{.\EOS\space}%
\providecommand \EOS [0]{\spacefactor3000\relax}%
\providecommand \BibitemShut  [1]{\csname bibitem#1\endcsname}%
\let\auto@bib@innerbib\@empty
\bibitem [{\citenamefont {Lindhard}\ \emph {et~al.}(1963)\citenamefont
  {Lindhard}, \citenamefont {Nielsen}, \citenamefont {Scharff},\ and\
  \citenamefont {Thomsen}}]{lindhard:1963}%
  \BibitemOpen
  \bibfield  {author} {\bibinfo {author} {\bibfnamefont {J.}~\bibnamefont
  {Lindhard}}, \bibinfo {author} {\bibfnamefont {V.}~\bibnamefont {Nielsen}},
  \bibinfo {author} {\bibfnamefont {M.}~\bibnamefont {Scharff}},\ and\ \bibinfo
  {author} {\bibfnamefont {P.}~\bibnamefont {Thomsen}},\ }\href@noop {}
  {\bibfield  {journal} {\bibinfo  {journal} {Kong.Dan.Vid.Sel.Mat.Fys.Med.}\
  }\textbf {\bibinfo {volume} {33}},\ \bibinfo {pages} {10} (\bibinfo {year}
  {1963})}\BibitemShut {NoStop}%
\bibitem [{\citenamefont {Dougherty}(1992)}]{Brian}%
  \BibitemOpen
  \bibfield  {author} {\bibinfo {author} {\bibfnamefont {B.~L.}\ \bibnamefont
  {Dougherty}},\ }\href {https://doi.org/10.1103/PhysRevA.45.2104} {\bibfield
  {journal} {\bibinfo  {journal} {Phys. Rev. A}\ }\textbf {\bibinfo {volume}
  {45}},\ \bibinfo {pages} {2104} (\bibinfo {year} {1992})}\BibitemShut
  {NoStop}%
\bibitem [{\citenamefont {Sattler}(1965)}]{SattlerDat}%
  \BibitemOpen
  \bibfield  {author} {\bibinfo {author} {\bibfnamefont {A.~R.}\ \bibnamefont
  {Sattler}},\ }\href@noop {} {\bibfield  {journal} {\bibinfo  {journal} {Phys.
  Rev}\ }\textbf {\bibinfo {volume} {138}},\ \bibinfo {pages} {1815} (\bibinfo
  {year} {1965})}\BibitemShut {NoStop}%
\bibitem [{\citenamefont {Zecher}\ \emph {et~al.}(1990)\citenamefont {Zecher},
  \citenamefont {Wang}, \citenamefont {Rapaport}, \citenamefont {Martoff},\
  and\ \citenamefont {Young}}]{Zech}%
  \BibitemOpen
  \bibfield  {author} {\bibinfo {author} {\bibfnamefont {P.}~\bibnamefont
  {Zecher}}, \bibinfo {author} {\bibfnamefont {D.}~\bibnamefont {Wang}},
  \bibinfo {author} {\bibfnamefont {J.}~\bibnamefont {Rapaport}}, \bibinfo
  {author} {\bibfnamefont {C.~J.}\ \bibnamefont {Martoff}},\ and\ \bibinfo
  {author} {\bibfnamefont {B.~A.}\ \bibnamefont {Young}},\ }\href
  {https://doi.org/10.1103/PhysRevA.41.4058} {\bibfield  {journal} {\bibinfo
  {journal} {Phys. Rev. A}\ }\textbf {\bibinfo {volume} {41}},\ \bibinfo
  {pages} {4058} (\bibinfo {year} {1990})}\BibitemShut {NoStop}%
\bibitem [{\citenamefont {Sarkis}\ \emph {et~al.}(2020)\citenamefont {Sarkis},
  \citenamefont {Aguilar-Arevalo},\ and\ \citenamefont {D'Olivo}}]{QFUNAM}%
  \BibitemOpen
  \bibfield  {author} {\bibinfo {author} {\bibfnamefont {Y.}~\bibnamefont
  {Sarkis}}, \bibinfo {author} {\bibfnamefont {A.}~\bibnamefont
  {Aguilar-Arevalo}},\ and\ \bibinfo {author} {\bibfnamefont {J.~C.}\
  \bibnamefont {D'Olivo}},\ }\href
  {https://doi.org/10.1103/PhysRevD.101.102001} {\bibfield  {journal} {\bibinfo
   {journal} {Phys. Rev. D}\ }\textbf {\bibinfo {volume} {101}},\ \bibinfo
  {pages} {102001} (\bibinfo {year} {2020})}\BibitemShut {NoStop}%
\bibitem [{\citenamefont {Chavarria}\ \emph {et~al.}(2016)\citenamefont
  {Chavarria} \emph {et~al.}}]{chavarria:2016}%
  \BibitemOpen
  \bibfield  {author} {\bibinfo {author} {\bibfnamefont {A.~E.}\ \bibnamefont
  {Chavarria}} \emph {et~al.},\ }\href
  {https://doi.org/10.1103/PhysRevD.94.082007} {\bibfield  {journal} {\bibinfo
  {journal} {Phys. Rev.}\ }\textbf {\bibinfo {volume} {D94}},\ \bibinfo {pages}
  {082007} (\bibinfo {year} {2016})},\ \Eprint
  {https://arxiv.org/abs/1608.00957} {arXiv:1608.00957 [astro-ph.IM]}
  \BibitemShut {NoStop}%
\bibitem [{\citenamefont {Izraelevitch}\ \emph {et~al.}(2017)\citenamefont
  {Izraelevitch} \emph {et~al.}}]{izraelevitch:2017}%
  \BibitemOpen
  \bibfield  {author} {\bibinfo {author} {\bibfnamefont {F.}~\bibnamefont
  {Izraelevitch}} \emph {et~al.},\ }\href
  {https://doi.org/10.1088/1748-0221/12/06/P06014} {\bibfield  {journal}
  {\bibinfo  {journal} {JINST}\ }\textbf {\bibinfo {volume} {12}}\bibfield
  {number} {\bibinfo  {number} { (06)},\ \bibinfo {pages} {P06014}},\ }\Eprint
  {https://arxiv.org/abs/1702.00873} {arXiv:1702.00873 [physics.ins-det]}
  \BibitemShut {NoStop}%
\bibitem [{\citenamefont {Frenkel}(1926)}]{Frenkel1926}%
  \BibitemOpen
  \bibfield  {author} {\bibinfo {author} {\bibfnamefont {J.}~\bibnamefont
  {Frenkel}},\ }\href {https://doi.org/10.1007/BF01379812} {\bibfield
  {journal} {\bibinfo  {journal} {Zeitschrift f{\"u}r Physik}\ }\textbf
  {\bibinfo {volume} {35}},\ \bibinfo {pages} {652} (\bibinfo {year}
  {1926})}\BibitemShut {NoStop}%
\bibitem [{\citenamefont {{J.Lindhard}}(1954)}]{Lindhard1954}%
  \BibitemOpen
  \bibfield  {author} {\bibinfo {author} {\bibnamefont {{J.Lindhard}}},\
  }\href@noop {} {\bibfield  {journal} {\bibinfo  {journal}
  {{Matematisk-fysiske Meddelelser}}\ }\textbf {\bibinfo {volume} {Vol: 28, No.
  8}} (\bibinfo {year} {1954})}\BibitemShut {NoStop}%
\bibitem [{\citenamefont {Sigmund}(2006)}]{PeterSigmundV1}%
  \BibitemOpen
  \bibfield  {author} {\bibinfo {author} {\bibfnamefont {P.}~\bibnamefont
  {Sigmund}},\ }\href@noop {} {\emph {\bibinfo {title} {Particle Penetration
  and Radiation Effects}}}\ (\bibinfo  {publisher} {Springer-Verlag Berlin
  Heidelberg},\ \bibinfo {year} {2006})\BibitemShut {NoStop}%
\bibitem [{\citenamefont {Migdal}(1941)}]{migdal1}%
  \BibitemOpen
  \bibfield  {author} {\bibinfo {author} {\bibfnamefont {A.~B.}\ \bibnamefont
  {Migdal}},\ }\href@noop {} {\bibfield  {journal} {\bibinfo  {journal} {J.
  Phys}\ }\textbf {\bibinfo {volume} {4}},\ \bibinfo {pages} {449} (\bibinfo
  {year} {1941})}\BibitemShut {NoStop}%
\bibitem [{\citenamefont {Baur}\ \emph {et~al.}(1983)\citenamefont {Baur},
  \citenamefont {Rosel},\ and\ \citenamefont {Trautmann}}]{migdal2}%
  \BibitemOpen
  \bibfield  {author} {\bibinfo {author} {\bibfnamefont {G.}~\bibnamefont
  {Baur}}, \bibinfo {author} {\bibfnamefont {F.}~\bibnamefont {Rosel}},\ and\
  \bibinfo {author} {\bibfnamefont {D.}~\bibnamefont {Trautmann}},\ }\href
  {https://doi.org/10.1088/0022-3700/16/14/006} {\bibfield  {journal} {\bibinfo
   {journal} {Journal of Physics B: Atomic and Molecular Physics}\ }\textbf
  {\bibinfo {volume} {16}},\ \bibinfo {pages} {L419} (\bibinfo {year}
  {1983})}\BibitemShut {NoStop}%
\bibitem [{\citenamefont {{F. Ziegler, J.P. Biersack, and U.
  Littmark}}(1985)}]{Ziegler}%
  \BibitemOpen
  \bibfield  {author} {\bibinfo {author} {\bibnamefont {{F. Ziegler, J.P.
  Biersack, and U. Littmark}}},\ }\href@noop {} {\bibfield  {journal} {\bibinfo
   {journal} {Pergamon Press New York}\ } (\bibinfo {year} {1985})}\BibitemShut
  {NoStop}%
\bibitem [{\citenamefont {Wilson}\ \emph {et~al.}(1977)\citenamefont {Wilson},
  \citenamefont {Haggmark},\ and\ \citenamefont {Biersack}}]{WilsonPot}%
  \BibitemOpen
  \bibfield  {author} {\bibinfo {author} {\bibfnamefont {W.~D.}\ \bibnamefont
  {Wilson}}, \bibinfo {author} {\bibfnamefont {L.~G.}\ \bibnamefont
  {Haggmark}},\ and\ \bibinfo {author} {\bibfnamefont {J.~P.}\ \bibnamefont
  {Biersack}},\ }\href {https://doi.org/10.1103/PhysRevB.15.2458} {\bibfield
  {journal} {\bibinfo  {journal} {Phys. Rev. B}\ }\textbf {\bibinfo {volume}
  {15}},\ \bibinfo {pages} {2458} (\bibinfo {year} {1977})}\BibitemShut
  {NoStop}%
\bibitem [{\citenamefont {Lindhard}\ and\ \citenamefont
  {Scharff}(1961)}]{lindhard1961}%
  \BibitemOpen
  \bibfield  {author} {\bibinfo {author} {\bibfnamefont {J.}~\bibnamefont
  {Lindhard}}\ and\ \bibinfo {author} {\bibfnamefont {M.}~\bibnamefont
  {Scharff}},\ }\href {https://doi.org/10.1103/PhysRev.124.128} {\bibfield
  {journal} {\bibinfo  {journal} {Phys. Rev.}\ }\textbf {\bibinfo {volume}
  {124}},\ \bibinfo {pages} {128} (\bibinfo {year} {1961})}\BibitemShut
  {NoStop}%
\bibitem [{\citenamefont {Semrad}(1986)}]{CoulombRepultionEffect}%
  \BibitemOpen
  \bibfield  {author} {\bibinfo {author} {\bibfnamefont {D.}~\bibnamefont
  {Semrad}},\ }\href {https://doi.org/10.1103/PhysRevA.33.1646} {\bibfield
  {journal} {\bibinfo  {journal} {Phys. Rev. A}\ }\textbf {\bibinfo {volume}
  {33}},\ \bibinfo {pages} {1646} (\bibinfo {year} {1986})}\BibitemShut
  {NoStop}%
\bibitem [{\citenamefont {Ovchinnikov}\ \emph {et~al.}(2004)\citenamefont
  {Ovchinnikov}, \citenamefont {Ogurtsov}, \citenamefont {Macek},\ and\
  \citenamefont {Gordeev}}]{Nonperturbative}%
  \BibitemOpen
  \bibfield  {author} {\bibinfo {author} {\bibfnamefont {S.}~\bibnamefont
  {Ovchinnikov}}, \bibinfo {author} {\bibfnamefont {G.}~\bibnamefont
  {Ogurtsov}}, \bibinfo {author} {\bibfnamefont {J.}~\bibnamefont {Macek}},\
  and\ \bibinfo {author} {\bibfnamefont {Y.}~\bibnamefont {Gordeev}},\ }\href
  {https://doi.org/https://doi.org/10.1016/j.physrep.2003.09.005} {\bibfield
  {journal} {\bibinfo  {journal} {Physics Reports}\ }\textbf {\bibinfo {volume}
  {389}},\ \bibinfo {pages} {119} (\bibinfo {year} {2004})}\BibitemShut
  {NoStop}%
\bibitem [{\citenamefont {Fano}\ and\ \citenamefont {Lichten}(1965)}]{Ufano}%
  \BibitemOpen
  \bibfield  {author} {\bibinfo {author} {\bibfnamefont {U.}~\bibnamefont
  {Fano}}\ and\ \bibinfo {author} {\bibfnamefont {W.}~\bibnamefont {Lichten}},\
  }\href {https://doi.org/10.1103/PhysRevLett.14.627} {\bibfield  {journal}
  {\bibinfo  {journal} {Phys. Rev. Lett.}\ }\textbf {\bibinfo {volume} {14}},\
  \bibinfo {pages} {627} (\bibinfo {year} {1965})}\BibitemShut {NoStop}%
\bibitem [{\citenamefont
  {Sigmund}(1982{\natexlab{a}})}]{SemiclassicalCrossSectionDer}%
  \BibitemOpen
  \bibfield  {author} {\bibinfo {author} {\bibfnamefont {P.}~\bibnamefont
  {Sigmund}},\ }\href {https://doi.org/10.1103/PhysRevA.26.2497} {\bibfield
  {journal} {\bibinfo  {journal} {Phys. Rev. A}\ }\textbf {\bibinfo {volume}
  {26}},\ \bibinfo {pages} {2497} (\bibinfo {year}
  {1982}{\natexlab{a}})}\BibitemShut {NoStop}%
\bibitem [{\citenamefont {Trubnikov}\ and\ \citenamefont
  {Yavlinskii}()}]{SemiclassicalCrossSectionDer2}%
  \BibitemOpen
  \bibfield  {author} {\bibinfo {author} {\bibfnamefont {B.}~\bibnamefont
  {Trubnikov}}\ and\ \bibinfo {author} {\bibnamefont {Yavlinskii}},\ }\href
  {http://www.jetp.ras.ru/cgi-bin/e/index/e/21/1/p167?a=list} {\bibfield
  {journal} {\bibinfo  {journal} {Zh. Eksp. Teor.Fiz.}\ }\textbf {\bibinfo
  {volume} {48}},\ \bibinfo {pages} {253}}\BibitemShut {NoStop}%
\bibitem [{\citenamefont {J.Lindhard}(1961)}]{lindhard:1961}%
  \BibitemOpen
  \bibfield  {author} {\bibinfo {author} {\bibnamefont {J.Lindhard}},\
  }\href@noop {} {\bibfield  {journal} {\bibinfo  {journal}
  {Kong.Dan.Vid.Sel.Mat.Fys.Med.}\ }\textbf {\bibinfo {volume} {28}} (\bibinfo
  {year} {1961})}\BibitemShut {NoStop}%
\bibitem [{\citenamefont {Tilinin}(1988)}]{Tilinin}%
  \BibitemOpen
  \bibfield  {author} {\bibinfo {author} {\bibfnamefont {I.~S.}\ \bibnamefont
  {Tilinin}},\ }\href {file:///tmp/mozilla_youssef0/e_067_08_1570-1.pdf}
  {\bibfield  {journal} {\bibinfo  {journal} {Zh. Eksp. Teor. Fiz.}\ }\textbf
  {\bibinfo {volume} {94}},\ \bibinfo {pages} {96} (\bibinfo {year}
  {1988})}\BibitemShut {NoStop}%
\bibitem [{\citenamefont {Sigmund}(1982{\natexlab{b}})}]{KineticTheorySe_Der}%
  \BibitemOpen
  \bibfield  {author} {\bibinfo {author} {\bibfnamefont {P.}~\bibnamefont
  {Sigmund}},\ }\href {https://doi.org/10.1103/PhysRevA.26.2497} {\bibfield
  {journal} {\bibinfo  {journal} {Phys. Rev. A}\ }\textbf {\bibinfo {volume}
  {26}},\ \bibinfo {pages} {2497} (\bibinfo {year}
  {1982}{\natexlab{b}})}\BibitemShut {NoStop}%
\bibitem [{\citenamefont {Kishinevsky}(1962)}]{Kishi}%
  \BibitemOpen
  \bibfield  {author} {\bibinfo {author} {\bibfnamefont {L.}~\bibnamefont
  {Kishinevsky}},\ }\href@noop {} {\bibfield  {journal} {\bibinfo  {journal}
  {Izv. Akad. Nauk SSSR}\ }\textbf {\bibinfo {volume} {26}},\ \bibinfo {pages}
  {1410} (\bibinfo {year} {1962})}\BibitemShut {NoStop}%
\bibitem [{\citenamefont {Firsov}(1959)}]{Firsov}%
  \BibitemOpen
  \bibfield  {author} {\bibinfo {author} {\bibfnamefont {Z.~E.}\ \bibnamefont
  {Firsov}, \bibfnamefont {O.B.}},\ }\href@noop {} {\bibfield  {journal}
  {\bibinfo  {journal} {Teor. Fiz.}\ }\textbf {\bibinfo {volume} {36}},\
  \bibinfo {pages} {1517} (\bibinfo {year} {1959})}\BibitemShut {NoStop}%
\bibitem [{\citenamefont {Robinson}(1974)}]{Kpar1}%
  \BibitemOpen
  \bibfield  {author} {\bibinfo {author} {\bibfnamefont {M.}~\bibnamefont
  {Robinson}},\ }\href@noop {} {\bibfield  {journal} {\bibinfo  {journal}
  {Third National Conference on Atomic Collisions with Solids, Kiev USRR}\ }
  (\bibinfo {year} {1974})}\BibitemShut {NoStop}%
\bibitem [{\citenamefont {{S. Oen}}\ and\ \citenamefont {{T.
  Robinson}}(1976)}]{Kpar2}%
  \BibitemOpen
  \bibfield  {author} {\bibinfo {author} {\bibfnamefont {O.}~\bibnamefont {{S.
  Oen}}}\ and\ \bibinfo {author} {\bibfnamefont {M.}~\bibnamefont {{T.
  Robinson}}},\ }\href
  {https://doi.org/https://doi.org/10.1016/0029-554X(76)90806-5} {\bibfield
  {journal} {\bibinfo  {journal} {Nuclear Instruments and Methods}\ }\textbf
  {\bibinfo {volume} {132}},\ \bibinfo {pages} {647} (\bibinfo {year}
  {1976})}\BibitemShut {NoStop}%
\bibitem [{\citenamefont {Littmark}(1976)}]{Kpar3}%
  \BibitemOpen
  \bibfield  {author} {\bibinfo {author} {\bibfnamefont {U.}~\bibnamefont
  {Littmark}},\ }\href@noop {} {\bibfield  {journal} {\bibinfo  {journal} {JPP
  9/21}\ } (\bibinfo {year} {Max-Planck Institut fur Plasmaphysik,
  1976})}\BibitemShut {NoStop}%
\bibitem [{\citenamefont {Fernández-Varea}\ and\ \citenamefont
  {Arista}(2014)}]{Arista}%
  \BibitemOpen
  \bibfield  {author} {\bibinfo {author} {\bibfnamefont {J.~M.}\ \bibnamefont
  {Fernández-Varea}}\ and\ \bibinfo {author} {\bibfnamefont {N.~R.}\
  \bibnamefont {Arista}},\ }\href
  {https://doi.org/https://doi.org/10.1016/j.radphyschem.2013.08.015}
  {\bibfield  {journal} {\bibinfo  {journal} {Radiation Physics and Chemistry}\
  }\textbf {\bibinfo {volume} {96}},\ \bibinfo {pages} {88} (\bibinfo {year}
  {2014})}\BibitemShut {NoStop}%
\bibitem [{\citenamefont {Sigmund}(2014)}]{PeterSigmundV2}%
  \BibitemOpen
  \bibfield  {author} {\bibinfo {author} {\bibfnamefont {P.}~\bibnamefont
  {Sigmund}},\ }\href@noop {} {\emph {\bibinfo {title} {Particle Penetration
  and Radiation Effects Volume 2}}}\ (\bibinfo  {publisher} {Springer
  International Publishing},\ \bibinfo {year} {2014})\BibitemShut {NoStop}%
\bibitem [{\citenamefont {Sigmund}(2008)}]{Sigmund2008Resume}%
  \BibitemOpen
  \bibfield  {author} {\bibinfo {author} {\bibfnamefont {P.}~\bibnamefont
  {Sigmund}},\ }\href {https://doi.org/10.3103/S1062873808050018} {\bibfield
  {journal} {\bibinfo  {journal} {Bulletin of the Russian Academy of Sciences:
  Physics}\ }\textbf {\bibinfo {volume} {72}},\ \bibinfo {pages} {569}
  (\bibinfo {year} {2008})}\BibitemShut {NoStop}%
\bibitem [{SeD(2000)}]{SeData1-ARSTILA2000473}%
  \BibitemOpen
  \bibfield  {journal} {\bibinfo  {journal} {Nuclear Instruments and Methods in
  Physics Research Section B: Beam Interactions with Materials and Atoms}\
  }\href {https://doi.org/https://doi.org/10.1016/S0168-583X(00)00050-1}
  {https://doi.org/10.1016/S0168-583X(00)00050-1} (\bibinfo {year}
  {2000})\BibitemShut {NoStop}%
\bibitem [{\citenamefont {N.Hecking}()}]{SeData2-N.Hecking}%
  \BibitemOpen
  \bibfield  {author} {\bibinfo {author} {\bibnamefont {N.Hecking}},\
  }\href@noop {} {\bibfield  {journal} {\bibinfo  {journal} {Nuclear
  Instruments and Methods in Physics Research Section B: Beam Interactions with
  Materials and Atoms}\ }\textbf {\bibinfo {volume} {59}},\ \bibinfo {pages}
  {619}}\BibitemShut {NoStop}%
\bibitem [{\citenamefont {Zhang}\ and\ \citenamefont
  {Weber}(2003)}]{SeData3-PhysRevB.68.235317}%
  \BibitemOpen
  \bibfield  {author} {\bibinfo {author} {\bibfnamefont {Y.}~\bibnamefont
  {Zhang}}\ and\ \bibinfo {author} {\bibfnamefont {W.~J.}\ \bibnamefont
  {Weber}},\ }\href {https://doi.org/10.1103/PhysRevB.68.235317} {\bibfield
  {journal} {\bibinfo  {journal} {Phys. Rev. B}\ }\textbf {\bibinfo {volume}
  {68}},\ \bibinfo {pages} {235317} (\bibinfo {year} {2003})}\BibitemShut
  {NoStop}%
\bibitem [{\citenamefont {Gervasoni}\ and\ \citenamefont
  {Cruz-Jiménez}(1996)}]{BohrStrip1}%
  \BibitemOpen
  \bibfield  {author} {\bibinfo {author} {\bibfnamefont {J.~L.}\ \bibnamefont
  {Gervasoni}}\ and\ \bibinfo {author} {\bibfnamefont {S.}~\bibnamefont
  {Cruz-Jiménez}},\ }\href
  {https://doi.org/https://doi.org/10.1016/0969-806X(96)00001-1} {\bibfield
  {journal} {\bibinfo  {journal} {Radiation Physics and Chemistry}\ }\textbf
  {\bibinfo {volume} {48}},\ \bibinfo {pages} {433} (\bibinfo {year}
  {1996})}\BibitemShut {NoStop}%
\bibitem [{\citenamefont {Gümüş}\ and\ \citenamefont {Önder
  Kabaday}()}]{BohrStrip2}%
  \BibitemOpen
  \bibfield  {author} {\bibinfo {author} {\bibfnamefont {H.}~\bibnamefont
  {Gümüş}}\ and\ \bibinfo {author} {\bibnamefont {Önder Kabaday}},\
  }\href@noop {} {\bibfield  {journal} {\bibinfo  {journal} {Journal of
  Physical Science and Application}\ }\textbf {\bibinfo {volume}
  {3}}}\BibitemShut {NoStop}%
\bibitem [{\citenamefont {Friedel}(1952)}]{Friedel}%
  \BibitemOpen
  \bibfield  {author} {\bibinfo {author} {\bibfnamefont {J.}~\bibnamefont
  {Friedel}},\ }\href {https://doi.org/10.1080/14786440208561086} {\bibfield
  {journal} {\bibinfo  {journal} {The London, Edinburgh, and Dublin
  Philosophical Magazine and Journal of Science}\ }\textbf {\bibinfo {volume}
  {43}},\ \bibinfo {pages} {153} (\bibinfo {year} {1952})},\ \Eprint
  {https://arxiv.org/abs/https://doi.org/10.1080/14786440208561086}
  {https://doi.org/10.1080/14786440208561086} \BibitemShut {NoStop}%
\bibitem [{\citenamefont {Peñalba}\ \emph {et~al.}(1992)\citenamefont
  {Peñalba}, \citenamefont {Arnau},\ and\ \citenamefont
  {Echenique}}]{ZOscilations}%
  \BibitemOpen
  \bibfield  {author} {\bibinfo {author} {\bibfnamefont {M.}~\bibnamefont
  {Peñalba}}, \bibinfo {author} {\bibfnamefont {A.}~\bibnamefont {Arnau}},\
  and\ \bibinfo {author} {\bibfnamefont {P.}~\bibnamefont {Echenique}},\ }\href
  {https://doi.org/https://doi.org/10.1016/0168-583X(92)95773-K} {\bibfield
  {journal} {\bibinfo  {journal} {Nuclear Instruments and Methods in Physics
  Research Section B: Beam Interactions with Materials and Atoms}\ }\textbf
  {\bibinfo {volume} {67}},\ \bibinfo {pages} {66} (\bibinfo {year}
  {1992})}\BibitemShut {NoStop}%
\bibitem [{\citenamefont {Holmstr\"om}\ \emph {et~al.}(2008)\citenamefont
  {Holmstr\"om}, \citenamefont {Kuronen},\ and\ \citenamefont
  {Nordlund}}]{FrenkelSi}%
  \BibitemOpen
  \bibfield  {author} {\bibinfo {author} {\bibfnamefont {E.}~\bibnamefont
  {Holmstr\"om}}, \bibinfo {author} {\bibfnamefont {A.}~\bibnamefont
  {Kuronen}},\ and\ \bibinfo {author} {\bibfnamefont {K.}~\bibnamefont
  {Nordlund}},\ }\href {https://doi.org/10.1103/PhysRevB.78.045202} {\bibfield
  {journal} {\bibinfo  {journal} {Phys. Rev. B}\ }\textbf {\bibinfo {volume}
  {78}},\ \bibinfo {pages} {045202} (\bibinfo {year} {2008})}\BibitemShut
  {NoStop}%
\bibitem [{\citenamefont {Ying-Tai}\ \emph {et~al.}(2002)\citenamefont
  {Ying-Tai}, \citenamefont {Qi-Ren},\ and\ \citenamefont
  {Chun-Yuan}}]{CrystalModelYing_Tai_2002}%
  \BibitemOpen
  \bibfield  {author} {\bibinfo {author} {\bibfnamefont {L.}~\bibnamefont
  {Ying-Tai}}, \bibinfo {author} {\bibfnamefont {Z.}~\bibnamefont {Qi-Ren}},\
  and\ \bibinfo {author} {\bibfnamefont {G.}~\bibnamefont {Chun-Yuan}},\ }\href
  {https://doi.org/10.1088/0253-6102/38/3/361} {\bibfield  {journal} {\bibinfo
  {journal} {Communications in Theoretical Physics}\ }\textbf {\bibinfo
  {volume} {38}},\ \bibinfo {pages} {361} (\bibinfo {year} {2002})}\BibitemShut
  {NoStop}%
\bibitem [{\citenamefont {Sarkis}\ \emph {et~al.}(2021)\citenamefont {Sarkis}
  \emph {et~al.}}]{MagCEVNS2021}%
  \BibitemOpen
  \bibfield  {author} {\bibinfo {author} {\bibfnamefont {Y.}~\bibnamefont
  {Sarkis}} \emph {et~al.},\ }\href {https://indi.to/DqCzz} {\bibinfo {title}
  {Magnificent \textsc{CEVNS} 2021}},\ \bibinfo {howpublished}
  {\url{https://indi.to/DqCzz}} (\bibinfo {year} {6 to 7 Oct.
  2021})\BibitemShut {NoStop}%
\bibitem [{\citenamefont {Gerbier}\ \emph {et~al.}(1990)\citenamefont {Gerbier}
  \emph {et~al.}}]{Gerbier}%
  \BibitemOpen
  \bibfield  {author} {\bibinfo {author} {\bibnamefont {Gerbier}} \emph
  {et~al.},\ }\href {https://doi.org/10.1103/PhysRevD.42.3211} {\bibfield
  {journal} {\bibinfo  {journal} {Phys. Rev. D}\ }\textbf {\bibinfo {volume}
  {42}},\ \bibinfo {pages} {3211} (\bibinfo {year} {1990})}\BibitemShut
  {NoStop}%
\bibitem [{\citenamefont {Agnese}\ \emph {et~al.}(2018)\citenamefont {Agnese}
  \emph {et~al.}}]{AGNESE201871}%
  \BibitemOpen
  \bibfield  {author} {\bibinfo {author} {\bibfnamefont {R.}~\bibnamefont
  {Agnese}} \emph {et~al.},\ }\href
  {https://doi.org/https://doi.org/10.1016/j.nima.2018.07.028} {\bibfield
  {journal} {\bibinfo  {journal} {Nuclear Instruments and Methods in Physics
  Research Section A: Accelerators, Spectrometers, Detectors and Associated
  Equipment}\ }\textbf {\bibinfo {volume} {905}},\ \bibinfo {pages} {71}
  (\bibinfo {year} {2018})}\BibitemShut {NoStop}%
\end{thebibliography}%
\bibliographystyle{apsrev4-2}
\end{document}